\documentclass[twocolumn,showpacs,preprintnumbers,amsmath,amssymb,prl]{revtex4}

\usepackage{graphicx}% Include figure files
\usepackage{dcolumn}% Align table columns on decimal point
\usepackage{bm}% bold math

\begin{document}

\preprint{IFPAN, IFT-UW}

\title{Tightening of knots in proteins}

\author{Joanna I. Su{\l}kowska$^1$, Piotr Su{\l}kowski$^2$, Piotr Szymczak$^2$ and Marek Cieplak$^1$}
 \affiliation{$^1$ Institute of Physics, Polish Academy of Sciences, \\
Al. Lotnik\'ow 32/46, 02-668 Warsaw, Poland \\
$^2$ Institute of Theoretical Physics, University of Warsaw, \\
ul. Ho{\.z}a 69, 00-681 Warsaw, Poland \\}

\date{\today}

\begin{abstract}

We perform theoretical studies of stretching of 20 proteins with knots within
a coarse grained model. The knot's ends are found to jump to well defined
sequential locations that are associated with sharp turns whereas in homopolymers
they diffuse around
and eventually slide off. The waiting times of the jumps
are increasingly stochastic as the temperature is raised. Larger knots do not
return to their native locations when a protein is released after stretching.

\end{abstract}

\pacs{87.15.Aa, 87.14.Ee, 87.15.La, 82.37.Gk, 87.10.+e}% PACS, the Physics and Astronomy
\maketitle

%\section{Introduction}

Time and again, objects of non-trivial topology turn out to be relevant in physics.
Polymers provide examples of such a relevance as they may acquire
topologically non-trivial configurations known as knots \cite{km1,mlg,stasiak}.
In DNA's -- polymers which are nearly homogeneous -- knots arise spontaneously and
abundantly \cite{Bao2003,Stella}.
In proteins, however, they are a rarity. Knots in the native states of proteins
were first discovered by Mansfield in 1994
\cite{mansfield}. Further research \cite{taylor,stat-knot,taylor1} and especially
a survey by Virnau et al. \cite{vmk}, has led to an identification of 273 examples
of proteins with knots which constitutes less than 1\% of the structures deposited
in the Protein Data Bank. The biological function of knots in proteins remains to
be elucidated, but it is likely that such shapes are not accidental.
It should also be noted  that
these 273 proteins correspond to only three different topologies denoted as
$3_{1}$ (the trefoil knot), $4_{1}$ and $5_{2}$ where the main integer indicates the
number of crossings and the subscript -- a particular shape.
(When identifying a knot, it is assumed implicitly that
the protein terminals are connected by an outside segment that transforms
a two-ended chain into a closed loop).

In this Letter, we explore the dynamical behavior of a knot when
a protein is stretched, for example by a tip of an atomic force microscope.
Experiments on knot-tightening have been
performed recently \cite{Alam_2002} for
the bovine carbonic anhydrase protein (coded 1v9e),
which was also studied within all-atom simulations \cite{Ikai}.
Our study is based on molecular dynamics simulations
in a coarse-grained model that represents a protein as a chain of
the C$^{\alpha}$ atoms with effective attractive
contact interactions \cite{Goabe,biophysical}.
In contrast to the all-atom simulations, a coarse grained approach
allows for a survey of many proteins, incorporation of much larger statistics,
slower rates of pulling, and extensive variation of parameters.

We observe that knot tightening process in a stretched protein is dominated
by jumps, i.e. sudden displacements of positions of knot's ends
along the sequence towards each other.
These jumps have definite lengths and together with the final
location of a tightened knot they are specified by a local geometry of a protein chain.
The larger the size of a knot, or its level of topological
complication, the larger the number of jumps is observed before its final tightening.
However, such jumps are not observed in the dynamics of knot motion
on stretched polymers. In this case, the motion is of a diffusive character
\cite{Bao2003,Metzler2006,polymers}.% with the knot mobility increasing towards
%the chain ends.

In order to define the knotted core, i.e.
a minimal segment of amino acids that can be identified as a knot,
we use the KMT algorithm \cite{km1,taylor}. It involves removing
the  C$^{\alpha}$ atoms, one at a time, as long as the backbone does not intersect
a triangle set by the atom under consideration and its two immediate sequential neighbors.
As a result of this procedure, two end points of the knot are identified.
%For a $3_1$ knot, they are separated by at least 25 amino acids. 
%(and 42 in the native state \cite{vmk}).
%The knot identification depends on the conformation and as the protein gets stretched,
%the end points of the knot may depin and come closer together.
The knot's ends depend on the conformation and, as the protein gets stretched, they may depin
and come closer together.
We have studied 18  proteins with the trefoil knot $3_1$ (1j85, 1o6d, 1dmx, 1jd0, 1j86,
1ipa, 1js1, 1k3r, 1kop, 1nxz, 1v9e, 1x7p, 1v2x, 1fug, 1vh0, 1zrj, 1hcb, 1keq)
and two $5_2$ proteins (2etl and 1xd3) \cite{vmk}.
We have found that once the knot shrinks from its native size, one
end of a knot invariably lands in a sharp turn of a protein backbone. Then it moves again
until a final position corresponding to the tightest knot is reached.
In most cases, such turns contain proline which stiffens a backbone through
a ring structure that forms a backbone angle $\sim 75 \deg$.
The second frequent knot-stopping turn contains glicyne 
(in 1o6d, 1fug, 1vho, 1zrj, 1keq, 1v9e -- the latter also has
a turn with proline) which, due to the lack of the side chain, leads to
strongly sinuous local conformations of the backbone.
In one case (1hcb), the knot-stopping turn involved alanine. In the
absence of a sharp turn in a protein backbone, the knot is stopped
at the beginning of a helix. It should be noted that proteins with knots
have a shorter effective end-to-end length available for stretching,
which is similar to the
case of proteins with covalent disulfide bonds
between cysteins (not present in the proteins considered here). However, there are also
important
differences between the two: disulfide bonds stay in place whereas knots may move.

The details of our modeling of stretching are described in refs. \cite{thermtit,survey}.
Native contacts are defined through heavy atom overlaps and are assigned
the Lennard-Jones potentials with an amplitude $\epsilon$ and length parameters tuned in
such a way
as to guarantee that the native conformation of a protein corresponds to the global minimum
of
potential energy. The remaining non-native contacts are repulsive.
We take $\epsilon/k_B= 900 K$, which correlates well with the experimental data on
protein unfolding %\cite{cieplakmarszalek}.
($T=k_BT/\epsilon \sim 0.3$ corresponds to the room temperature).
Unlike Wallin et al. \cite{shak} who consider {\em folding}
of protein 1j85, we do not need to introduce additional non-native attractive contacts
leading to a knot formation, since our configurations are already knotted.

%A local backbone stiffness is maintained through a chirality potential \cite{survey}.
The presence of a solvent is mimicked by velocity dependent friction
and fluctuational forces corresponding to a temperature $T$.
The stretching was accomplished by attaching the protein to a pulling spring which
moves with the velocity $v_p$ of 0.005 {\AA}/ns.
Our approach and its variants has passed many benchmark tests
for protein stretching and agrees favorably with the experimental results \cite{survey}
and all-atom molecular dynamics simulations \cite{paci}.
In particular, our model predicts existence of three peaks in the force--displacement
curves (at 130, 370, and 490 {\AA}) for 1v9e as found in experiment \cite{Alam_2002,Ikai}
and all-atom simulations \cite{Ikai} and a similar order of contact breaking events.

\begin{figure}[htb]
\begin{center}
\includegraphics[width=0.32\textwidth]{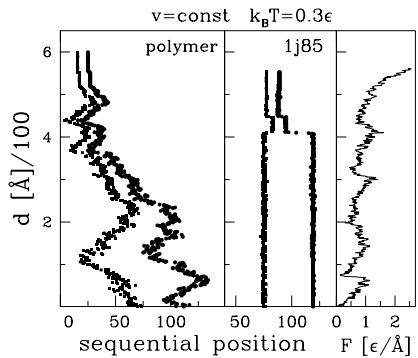}
\caption{Motion and tightening of a knot on a homopolymer (left)
and on protein 1j85 (middle) during stretching with constant velocity.
Squares and circles indicate positions of the ends of the knot along the chain.
Knots typically slide off homopolymeric chains. Here, however, we have chosen
an example in which a knot tightens close to one end of the chain.
In contrast, knots in proteins always tighten in a specific position inside its initial
configuration, after making a series of jumps.
Each jump corresponds to a definite force peak
in the force--displacement curve shown in the panel on the right hand side. } \label{fig-
levy}
\end{center}
\end{figure}

In order to represent motion of the ends of a knot we use diagrams such as one shown in the
middle panel of Figure 1. The panel corresponds to protein 1j85 which contains $N$=156
amino acids that make a simple trefoil knot with the ends set at amino acids
numbered $n_1$=75 and $n_2$=119 in the native state.
The diagram shows what happens to the values of $n_i$ when the protein gets
stretched as the pulling tip moves by a distance $d$ and the corresponding
force ($F$) -- displacement ($d$) curve for a protein (the right panel in Figure 1).
It is seen that despite the presence of several force peaks
the ends of the knot stay put for most of the stretching trajectory.
However, at the final force peak, i.e. around $d$=400 {\AA},
both ends jump towards each other along the sequence and then
undergo another jump about 50 {\AA} later. This jumpy behavior is not found when the
protein is heated up or replaced by a homopolymer with purely repulsive
contact interactions (the left panel in Figure 1). In the homopolymeric case,
we start with the native conformation of a protein, but remove attractive contacts.
Another possibility of observing homopolymer-like
behavior in a protein is to increase the temperature of the system above that of the
specific-heat maximum. In hompolymer, the positions of the knot ends diffuse around
and, particularly in the initial stages, the distance between them may increase considerably
which corresponds to swelling of the knot. Eventually, however, they come closer together
but remain mobile and, in most cases, slide off the polymer chain.
These results agree with earlier studies on the dynamics of knots in polymers and DNA, in
which the diffusive character of
knot motion was analyzed both experimentally \cite{Bao2003} and theoretically \cite{
Metzler2006,polymers}.
%The main reason of such a contrasting behavior between polymers
%and proteins is the presence of a complex network of bonds between aminoacids in the
%latter.
%We confirmed that by simulating the proteins with all attractive contact interactions
%removed,
%in which case a polymer behavior is indeed seen.

Both for the homopolymer and the protein, the motion of the knot's ends
depends on the particular trajectory even if the $F(d)$ curves look nearly the same.
In particular, the ends may sometimes depin on an earlier force peak.
%Similar knot-end wandering diagrams are obtained when  a protein was stretched at constant
%force instead of constant velocity.
The stretching process affects the knotted core of a protein much
less than the outside region and thus leaves the geometry inside
the knotted core and its secondary structures nearly native-like.
For instance, a well tightened knot in 1o6d contains an entire $\alpha$-helix
in its nearly native conformation.
%For instance, in 1o6d the almost tightened knot contains entire
%$\alpha$-helix, grossly intact with respect to its native configuration.

The description of a knot dynamics is reduced and involves only the movement 
of its end points $n_i$ along the sequence. We have found, however, that the real space
distances between the residues in the knotted core turn out to be mostly unchanged in between the knot jumps
and undergo rapid changes as the knot ends jump. This indicates the existence of a coupling of
the real space dynamics of a knot to its motion in the sequence space.

\begin{figure}[htb]
\begin{center}
\includegraphics[width=0.35\textwidth]{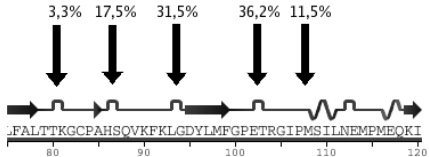}
\caption{The ends of a knot in 1j85 protein in the native state are located at amino
acids $n_1=75$ and $n_2=119$. In a tightened configuration, the ends of the knot are
located between $n_1$ and $n_2$, with one end either in a sharp turn or at the end
of a helix. The arrows indicate these characteristic places.
The numbers show percentages of situations (based on 700 trajectories)
in which a knot's end is pinned at the feature after moving from the native state.
The innermost features correspond to the tightest knot.
} \label{fig-1j85-turns}
\end{center}
\end{figure}

The final and metastable locations of the knot ends coincide with the sharp turns
in the protein backbone (and/or the endpoints of a helix), as seen in Figures
2 and 3. The stopping points correspond to the deep
local minima of the angle $\theta$ between every second vector along the $C_{\alpha}$
backbone (i.e between the vector $C_{\alpha,i}—C_{\alpha,i+1}$ and
$C_{\alpha,i+2}—C_{\alpha,i+3}$), which coincides with Kuntz's criterion \cite{kuntz}
for detection of turns (and is also satisfied at the end points of a helix).
Such turns are usually stabilized by hydrogen bonds and are thus harder to break.
At high temperatures ($kT > 0.5 \epsilon$),
%sharp angled conformations may arise entropically
the motion of a knot gradually becomes less predictable, and the final position
of the knot ends is no longer always connected to the turn in the native structure.
Additionally, the knot may wander outside the initial knotted core. Finally, for
$kT \gg \epsilon$, a homopolymeric behavior is observed, with the knot freely diffusing
along the backbone.

\begin{figure}[htb]
\begin{center}
\includegraphics[width=0.25\textwidth]{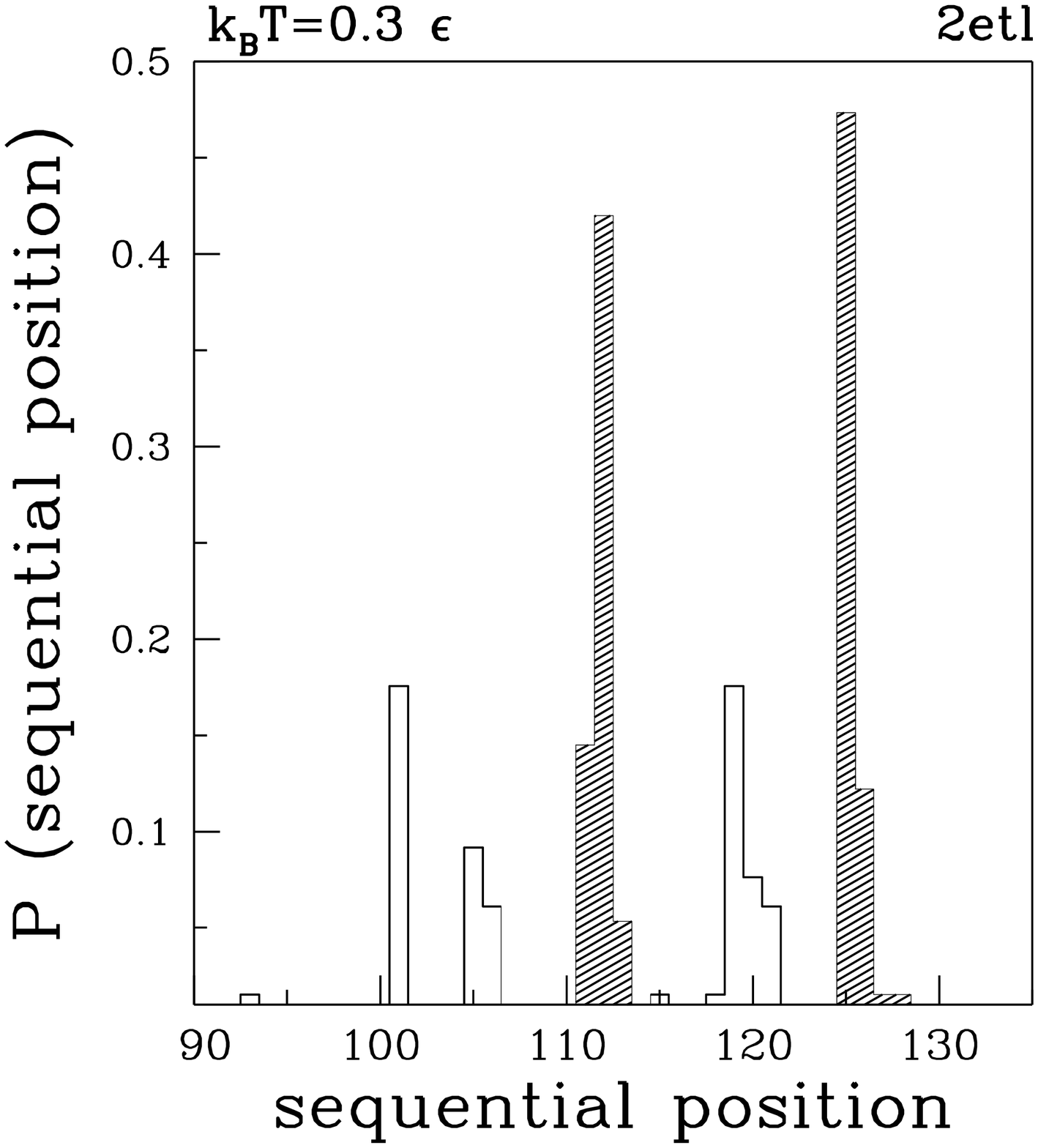}
\includegraphics[width=0.2\textwidth]{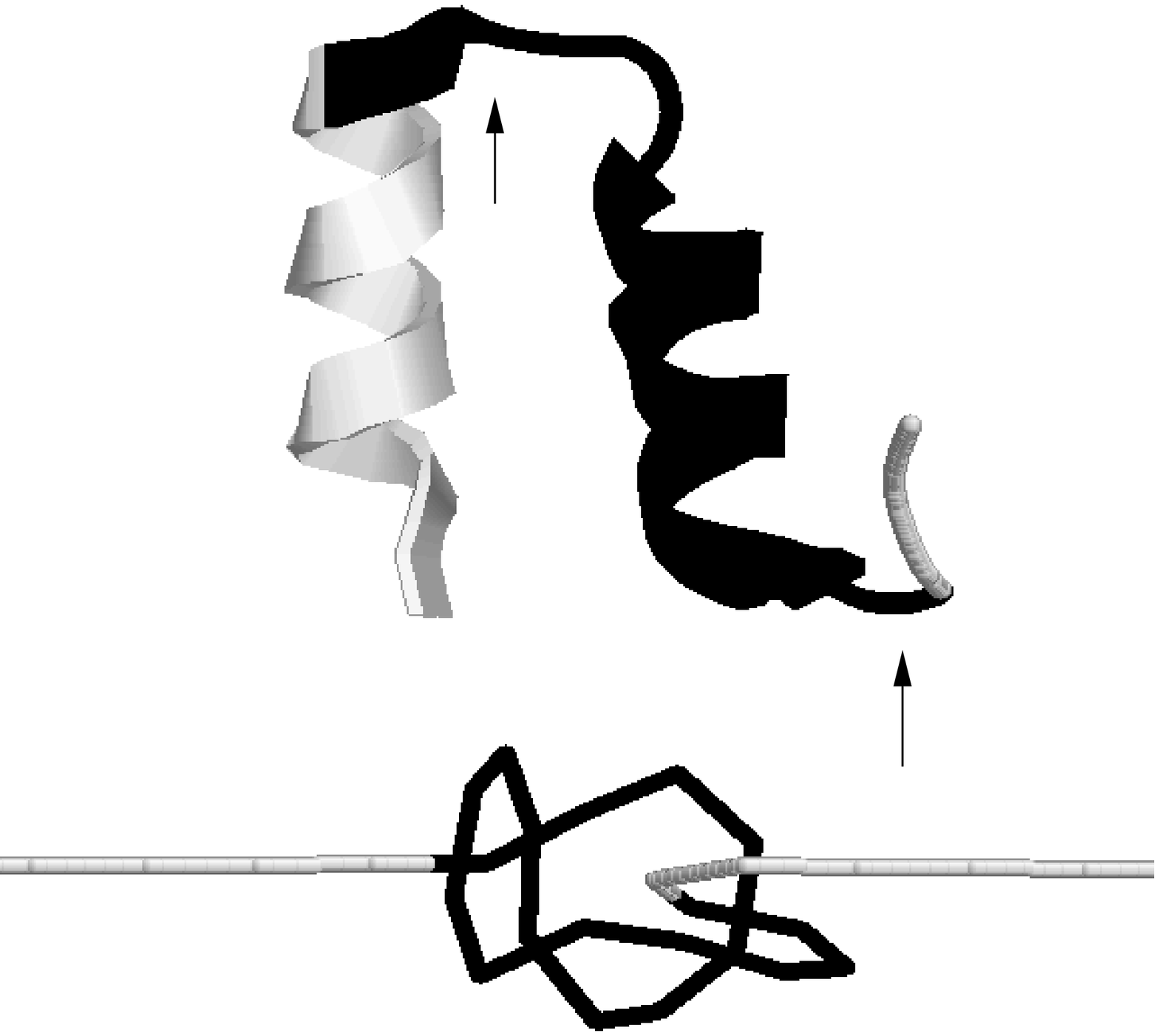}
\caption{ The preferred final locations of knot's ends in 2etl protein found in 700
trajectories. The darker peaks indicate to the most likely outcome:
$n_1\sim 111$ (the end of a helix) and $n_2\sim 126$ (a turn). The corresponding tightened
knot conformation
is shown on the bottom right. The relevant sequential segment is shown
on the top right in the native conformation where the arrows indicate
the values of $n_i$. The less probable outcomes are shown by the
lighter peaks. Here $n_1\sim 101$ or $106$ whereas $n_2\sim 119$.
} \label{fig-2etl-tie}
\end{center}
\end{figure}

A protein typically contains several sharp turns in the native state. Thus there
are several pinning centers on which the knot's ends may settle during
stretching. This is illustrated in Figure \ref{fig-1j85-turns} for 1j85 protein.
Another example is given in Figure \ref{fig-2etl-tie}
for the 2etl protein which supports a $5_2$ knot spanning 174
(out of all 223) sites in the native state.
In this case, there are two characteristic pinning centers
leading to the final knot tightening either between sites 110--126 or 101--119
for a range of temperatures.
%Note that the former site corresponds to a tighter structure.
It should be noted that the preference for a knot to begin or end on a turn
does not appear to apply to the native conformation. It arises only during stretching.

\begin{figure}[htb]
\begin{center}
\includegraphics[width=0.32\textwidth]{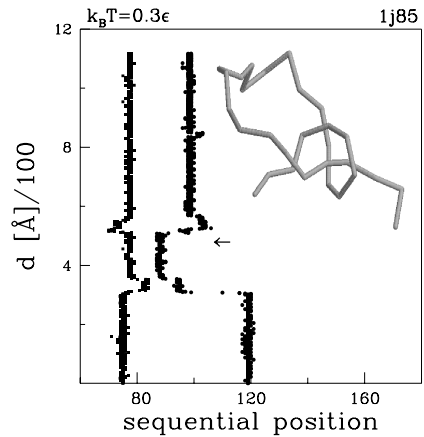}
\caption{After terminating the pulling process at $d=500$[A] (indicated by arrow)
1j85 returns to its native state in most cases.
However, sometimes it ends up in a metastable state as shown on the right.}
\label{fig-metastab}
\end{center}
\end{figure}

In addition to the simple stretching (whether at a constant speed or at a constant
force), we have also studied processes in which one pulls a protein to a certain
extension and then releases it abruptly. If the stretching stage lasts sufficiently long
(so that several force peaks are observed
and the knot gets tightened substantially)
then the protein misfolds on releasing
and the knot ends continue to reside at the metastable locations.
We have observed such irreversibility effects in 2etl, 1vho, and 1v2x
and in 80\% of trajectories for 1o6d.
% with knots size respectivly (174, 51, 51 and 42 ).
However, apart from a few trajectories (such as the one shown in Figure
\ref{fig-metastab}), the knot in protein 1j85 is usually found to return to its native
location. The different behavior of 1j85 compared to the other four proteins
may be due to the fact that 1j85 easily unfolds (and unties itself)
through heating \cite{shak} as well in equilibrium condition in the experiment \cite{mallam}
and is thus less stable.

\begin{figure}[htb]
\begin{center}
\includegraphics[width=0.4\textwidth]{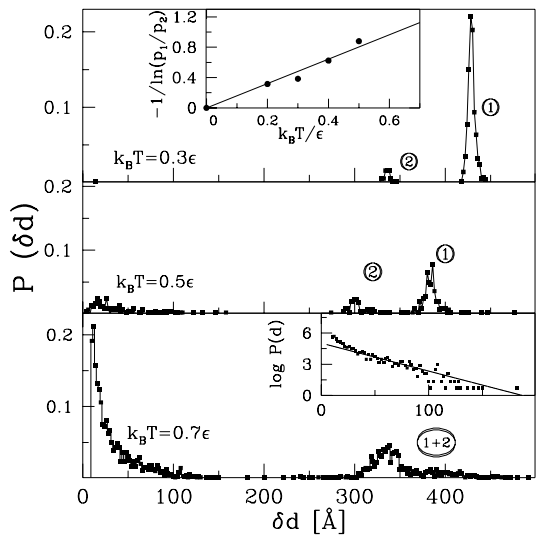}
\caption{Distribution of waiting distances $\delta d$ of the left end of a knot in protein
  1j85 at various temperatures $T$. Pathways 1 and 2 are indicated by symbols
in circles. The panels are explained in the text.}
\label{fig-temps}
\end{center}
\end{figure}

We now consider a distribution of waiting times, $\delta t$, between the jumps.
In fact, it is convenient to measure these times in terms of a respective displacement
of the pulling tip $\delta d = v_{p}\delta t$ 
(in addition, pulling distances corresponding to jumps are only weakly sensitive to the choice of $v_p$)
%(in particular, if the pulling velocity
%$v_{p}$ is changed, the times corresponding to unzipping of subsequent structures from the
%protein change accordingly but the corresponding $d$ values remain constant).
At $T$=0, the process
is deterministic, lasts for a relatively long time, with $\delta d$ reaching 400 {\AA} before
the first (and only) jump is made. At the time of the jump, the knotted core constitutes the
only portion of the original protein structure that has not been unfolded yet.
This unfolding route is denoted as pathway 1 and corresponds to the
rightmost peak in the top and middle panels of Figure \ref{fig-temps}.
As the temperature is increased an alternative pathway 2 becomes
stochastically available. In this pathway the knot is tightened at $\delta d\sim 340$ {\AA},
which is before the protein gets fully unfolded. 
%The remaining part of the protein structure
%consists of the first and third $\beta$ strands linked by hydrogen bonds, which will
%subsequently be broken.
% This is a metastable configuration for the knot.
The ratio of probabilities of choosing these pathways can be then described as
\begin{equation}
\frac{p_1}{p_2} = \exp\,\Big(-\frac{\Delta F}{k_B T}\Big),   \label{prob-temp}
\end{equation}
where $\Delta F$ is the free energy barrier associated with the transition
between pathway 1 and pathway 2.
The data points shown in the inset of the top panel of Figure \ref{fig-temps} suggest
$k_B/\Delta F \approx 1.6$.
As the temperature increases, the jumps on each pathway gets shorter and
are usually followed by another jump with
much shorter jumping distance
($d<100$ \AA$ $ in the middle and bottom panel).
Above $k_BT/\epsilon=0.5$ the peaks corresponding to pathways 1 and 2
merge. At this stage, the short distance part of the distribution may be
approximated by the exponential distribution
$P(d)= \alpha^{-1}\exp(\alpha d)$, as shown in the bottom panel for $k_BT/\epsilon=0.7$.
In the inset in the bottom panel $\textrm{log}\,P(d)$ is fitted to a line
whose slope yields $\alpha \approx$ -0.027.
%\aprox

In summary, we have found that the process of knot tightening in proteins is
qualitatively distinct from that occurring in homopolymers.
The proteinic knots shrink in size and one of their ends gets pinned on
a sharp turn. The movement of knot ends in the protein
along the sequence is characterized by sudden jumps,
whereas in polymers knots perform a diffusive motion and, in most cases, slide off the
chain. It would be interesting to devise stretching experiments that would monitor
knot tightening and end-jumping in proteins, analogous to those reported for
nucleic acids~\cite{Bao2003}.

We appreciate useful comments of R. Kutner and R. Stolarski.
This work was funded by the MNiSW grant N202 021 31/0739.

\end{document}